\documentclass[a4paper,11pt]{article}
\usepackage{pos}
\usepackage{float}
\usepackage{multirow}
\usepackage{subfigure}
\usepackage{braket}
\usepackage{tikz}

\newcommand{\beq}{\begin{equation}}
\newcommand{\eeq}{\end{equation}} 
\newcommand{\bea}{\begin{eqnarray}}
\newcommand{\eea}{\end{eqnarray}} 
\newcommand{\G}{\Gamma}
\newcommand\csg[1][2]{SU(#1)_L\times SU(#1)_R}
\newcommand{\AAS}{U(1)_A}
\newcommand\CS[1][2]{SU(#1)_{CS}}
\newcommand{\mv}{~\rm{MeV}}
\newcommand{\ggv}{~\rm{GeV}}
\newcommand{\Oop}{\mathcal{O}}

\newcommand{\q}{\overline{q}}

\newcommand{\dg}{^\dagger}
\newcommand{\cng}[1][\mu]{\gamma_#1}

\newcommand{\gt}{\cng[0]}
\newcommand{\gx}{\cng[1]}
\newcommand{\gy}{\cng[2]}
\newcommand{\gz}{\cng[3]}
\newcommand{\gv}{\cng[5]}
\newcommand{\ra}{\rightarrow}

\newcommand\xoh[1]{\frac{#1}{2}}
\newcommand{\oh}{\xoh{1}}
\newcommand\brb[1]{\bigg(#1 \bigg)}

\newcommand{\sgn}{\mbox{sgn}}

\title{Study of symmetries in finite temperature $N_f=2$ QCD with M\"obius Domain Wall Fermions}

\author*[a]{David Ward}
\author[b]{Sinya Aoki}
\author[c]{Yasumichi Aoki}
\author[a]{Hidenori Fukaya}
\author[d,e]{Shoji Hashimoto}
\author[c]{Issaku Kanamori}
\author[d,e,f]{Takashi Kaneko}
\author[c]{Jishnu Goswami}
\author[g]{Yu Zhang}
\onbehalf{(JLQCD Collaboration)}

\affiliation[a]{Department of Physics, Osaka University,\\  Toyonaka, Osaka 560-0043, Japan}

\affiliation[b]{Center for Gravitational Physics, Yukawa Institute for Theoretical Physics, Kyoto University,\\  Kyoto 650-0047, Japan}
  
\affiliation[c]{RIKEN Center for Computational Physics(Riken CCS),\\  Kobe 650-0047, Japan}
  
\affiliation[d]{KEK Theory center, High Energy Accelerator Research Organization(KEK),\\ Tsukuba 305-0801, Japan}

\affiliation[e]{School of High Energy Accelerator Science, Graduate University for Advanced Study(SOKENDAI),\\ Tsukuba 305-0801, Japan}

\affiliation[f]{Kobayashi-Maskawa Institute for the Origin of Particles and the Universe, Nagoya University,\\ Aichi 464-8603, Japan}

\affiliation[g]{Faculty of Physics, Bielefeld University, \\ Universitätsstraße 4, 33615 Bielefeld, Germany}

\renewcommand{\logo}{\relax}
\renewcommand{\hookAfterAbstract}{%
   \par\bigskip
   \textsc{Preprint Number}:
   \textsc{OU-HET-1249}  
   \textsc{KEK-CP-0406}
   \textsc{arXiv}:\href{https://arxiv.org/abs/2412.06574}{\color{black}{2412.06574}}
}

\emailAdd{ward@het.phys.sci.osaka-u.ac.jp}

\abstract{We report on the ongoing study of symmetry of $N_f=2$ QCD around the critical temperature. Our simulations of $N_f = 2$ QCD employ the M\"obius domain-wall fermion action with residual mass $\sim 1\mv$ or less, maintaining a good chiral symmetry. Using the screening masses from the two point spatial correlators we compare the mass difference between channels connected through various symmetry transformations. Our analysis focuses on restoration of the $SU(2)_L\times SU(2)_R$ as well as anomalously broken axial $U(1)_A$. We also present additional study of a potential $SU(2)_{CS}$ symmetry which may emerge at sufficiently high temperatures.}

\FullConference{The 41st International Symposium on Lattice Field Theory (Lattice 2024)\\
July 28th - August 3rd, 2024\\
University of Liverpool\\}

\begin{document}
\maketitle

\section{Introduction}
\label{sec:Intro}
Chiral symmetry is key to determining the phases of quantum chromodynamics(QCD). At low temperatures, spontaneous breaking of chiral symmetry is the origin of hadronic masses. Whereas at high temperatures, such as those in the early universe, restoration of chiral symmetry impacts the fundamental properties of the quark gluon plasma. It is, therefore, important to understand the chiral phase transition not only in particle physics, but also in both nuclear physics and cosmology.

The QCD Lagrangian is invariant under the $\csg[N_f]$ and $\AAS$ chiral symmetries in the massless limit. Because the axial $\AAS$ symmetry is broken by quantum anomaly at all energies, as a result it is natural to assume to play little role in the phase transition. Therefore, the $\csg[N_f]$ symmetry sector is considered the most relevant to describe the chiral phase transition.

It has been argued that the chiral condensate is not simply an order parameter for chiral symmetry transitions, but may in fact have a relationship to the axial anomaly $\AAS$ \cite{Aoki:2012yj, Aoki:2024uvl, Cossu:2013uua}. If the $\AAS$ symmetry is effectively restored at the critical temperature, then the order and universality of the phase transition may change \cite{Pisarski:1983ms}.

Preservation of chiral symmetry by operators in the lattice quark action is essential to the study of spontaneous restoration/breaking of chiral symmetry at high temperatures. In previous simulations, however, it has been difficult to preserve chiral symmetry and so non-chiral fermions have often been employed. Wilson fermions break the $\csg[2]$ to a remnant vector-like $SU(2)$; staggered fermions, while successful in many studies, break $\csg[2]$ to a remnant $U(1)_V\times U(1)_{A'}$ subgroup symmetry. Therefore, study of chiral symmetry using methods such as overlap or domain wall is attractive to produce more theoretically clean evidence of the chiral transition behaviors as the $\csg[2]$ is preserved.

In this work, we simulate $N_f=2$ QCD using M\"obius Domain Wall Fermions, with a residual mass $\sim 0.1\mv$. While some previous results at higher temperatures have already been published \cite{Suzuki:2019vzy,Rohrhofer:2019yko,Rohrhofer:2019qwq,Aoki:2020noz}, this most recent work covers previous temperatures and adds two new ensembles at $T=147\mv(0.9T_c)$ and $165\mv(T_c)$ allowing us to study symmetries around the critical temperature. To examine the symmetries of the light mesons we use the screening mass extracted from the spatial two-point meson correlation function in the isospin triplet channel.

In addition to the well known symmetries of QCD, we are interested in studying emergent symmetry structures which are not seen in the original QCD Lagrangian\cite{Glozman:2015qzf,Glozman:2017dfd,Glozman:2022lda,Glozman:2016swy}. In recent literature \cite{Rohrhofer:2019saj, Rohrhofer:2018xpn, Suzuki:2020rla} emergent higher symmetries of QCD e.g. $\CS[N_f]$, $SU(2N_f)$ have been increasingly discussed, which we have addressed in previous numerical simulations \cite{Rohrhofer:2019qwq,Zhang:2022kzb,Ward:2024tdm,Bala:2023iqu,Chiu:2023hnm,Chiu:2024jyz} and will explore in this work as well.

\section{Spatial Mesonic Correlators}
\label{sec:SpMesons}
Let us consider the quark isospin triplet bilinear operators 
\beq
\Oop_\G(x)=\q (x)(\G\otimes \xoh{\vec{\tau}})q(x) \label{eqn:Oopdef},
\eeq
where $\tau^a$ is an element of the generators of $SU(2)$, $\G$ denotes the combination of Dirac $\gamma$ listed in Table. \ref{tab:corrops}. Using the correlation functions of the $\Oop_\G(x)$ operators we examine symmetries at high temperature using the screening mass spectrum. Specifically we make use of the two point correlator along the $z$-axis,
\beq\label{eqn:zdspatialcorr}
C_\G(n_z)=\sum_{n_t,n_x,n_y} \braket{\Oop_\G(n_x,n_y,n_z,n_t)\Oop\dg_\G(0,0,0,0)}
\eeq
where $n_t$ is the Euclidean time direction and we specify the correlator channel with $\G$. Assuming confinement in the spatial direction of QCD, the correlation function at long distances is expected to exponentially decay following $C_\G(z)\sim e^{-M_{\G}z}$, where $M_\G$ is the screening mass of the lowest energy state in the $\G$ channel.\par

\begin{table}[ht!]
   \centering
   \begin{tabular}{c c c c c}
         \hline
         $\Gamma$ & Reference Name & Abbr. & \multicolumn{2}{c}{Symmetry Correspondences}\\
         \hline
         $\mathbb{I}$ & Scalar & S & \multirow{2}{*}{$\bigg\}U(1)_A$}\\
         $\gv$ & Psuedo Scalar & PS\\
         $\cng[1],\cng[2]$ & Vector & $\mathbf{V}$ & \multirow{2}{*}{$\bigg\}\csg$} & \multirow{4}{*}{$\bigg\} \CS?$ }\\
         $\cng[1]\gv,\cng[2]\gv$ & Axial Vector & $\mathbf{A}$\\
         $\cng[4]\gz$ & Tensor &  $\mathbf{T}_t$ & \multirow{2}{*}{$\bigg\}U(1)_A$} \\
         $\cng[4]\gz\gv$ & Axial Tensor &  $\mathbf{X}_t$\\
         \hline
   \end{tabular}
   \caption{The six channels of interest in this study and connecting symmetry transformations.}
   \label{tab:corrops}
\end{table}

\section{Emergence of $SU(2)_{CS}$}
For $T\ra \infty$ the free quark propagator with mass $m$ appears to exhibit an additional set of symmetries. Let us consider a quark propagator which has momenta $(p_x, p_y)$ perpendicular to the $z$-axis:
\beq
\braket{\q(z)q(0)}(p_1,p_2)=\sum_{p_0}\int_{-\infty}^\infty \frac{dp_3}{2\pi}\dfrac{m-(i\gt p_0+i\gamma_ip_i)}{p_0^2+p_i^2 +m^2}e^{ip_3z},
\eeq
where we have decomposed the Euclidean time direction into the Matsubara frequencies $p_0=2\pi T(n+\oh)$ and labelled $x,y$ and $z$ momenta with numerical indicies. The $p_3$ integral has a pole at $-ip_3=E=\sqrt{p_0^2+m^2+p_1^2+p_2^2}$ resulting in, 
\beq
\braket{\q(z)q(0)}(p_1,p_2)=\sum_{p_0}\dfrac{m+\gz E-(i\gt p_0+i\gx p_1+i\gy p_2)}{2E}e^{-Ez}.
\eeq
In the $T\ra \infty$ limit we can take $T\gg m^2+p_1^2+p_2^2$ and expand the correlator in terms of $1/T$:
\beq
\braket{\q(z)q(0)}=\sum_{p_0=\pm \pi T}\gz\xoh{1+i \sgn(p_0)\gt\gz}e^{-\pi Tz}+O(1/T).
\eeq
Here we have approximated the summation over $p_0$ to be the lowest Matsubara modes with $M=\pm \pi T$ only. This form of the correlator is invariant under an additional set of transformations:
\beq
\begin{split}
q(x)&\rightarrow \exp\brb{{\xoh{i}\Sigma_a\theta_a}}q(x)\\
\q (x)&\rightarrow \q(x)\gt \exp\brb{-\xoh{i}\Sigma_a\theta_a}\gt,
\end{split}
\eeq
where
$$\Sigma_k = \begin{bmatrix}
\cng[k]\\
i\cng[k]\gv\\
\gv
\end{bmatrix},$$
which is the so-called $\CS$ chiral-spin symmetry, explored in \cite{Glozman:2016swy,Glozman:2015qzf,Rohrhofer:2019saj, Rohrhofer:2019qwq,Bala:2023iqu}.\par 
For the free quark case shown above, the chiral-spin symmetry is approximate and emerges in the high temperature limit. However, when gluonic interactions are introduced the quark propagator has a nontrivial form which cannot be analyzed akin to the free quark form, yet it may possibly transform under the chiral-spin symmetry. Other work \cite{Rohrhofer:2019qwq,DallaBrida:2021ddx,Glozman:2017dfd} has shown evidence for an emergent $\CS$ symmetry at temperatures $\sim 1.1T_c$ and above. While it has also been argued this symmetry is over taken by interactions with the gluonic medium\cite{Chiu:2023hnm,Chiu:2024jyz} at very high temperatures.\par 
Although there are estimates  presented for this symmetry, it remains unclear if such a symmetry is emergent near $T_c$ and whether this symmetry is preserved into a higher temperature regime. For the range of temperatures in our study we also consider analysis of the channels which are related by $\CS$ symmetry transformation as a further examination of the $\CS$ around $T_c$.

\begin{figure}[t!]
\centering
\includegraphics[scale=1.2]{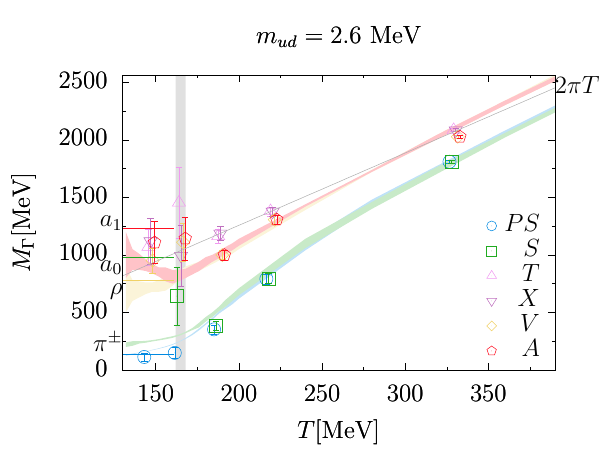}
\caption{For the lightest mass quark ensembles the $N_f=2$ screening mass data plotted with respect to temperature, shown as points. The shaded bands are $N_f=2+1$ data from HotQCD in \cite{Bazavov:2019www}.\label{fig:mch-nf2-0010}}
\end{figure}

\section{Numerical Results}
\label{sec:Simulations}
Our simulations of $N_f=2$ flavor QCD employ the tree-level improved Symanzik gauge action, as well as M\"obius domain wall fermion action with three steps of stout smearing. For all ensembles $\beta=4.30$ with a lattice cutoff set to $a^{-1}=2.463\ggv$. The residual mass of the domain-wall quark is reduced to $<0.1\mv$. For the highest temperature ensembles the aspect ratio is 4:1 and we do not expect finite size effects on the screening masses. For temperatures $T\sim 147\mv$ and $T\sim 165\mv$ lattices with spatial extent $L=36$ and $L=32$ respectively may not be enough to control the long range correlation effects for the spatial correlation functions and so we also consider additional lattices with spatial extents of $L=48$ and $L=40$ respectively. For all new ensemble data we considered four values of $am=0.0010,0.0025,0.00375,0.0050$ for up and down quark masses with the lightest value $am=0.0010$ corresponding to a mass of $2.6\mv$ below the physical point for the lightest quarks.
The value for the screening mass is extracted from fitting lattice data to the standard form
\beq\label{eqn:coshfits}
C(z)=A\cosh(m[z-L/2]).
\eeq
For channels with noisy fits such as the $S$-channel, and in particular the lowest two temperatures in our study we omit the fit values as both the correlator and the effective mass curves are too unstable to produce a reliable fit value.

\begin{figure}[ht]
\centering
\includegraphics[scale=1.2]{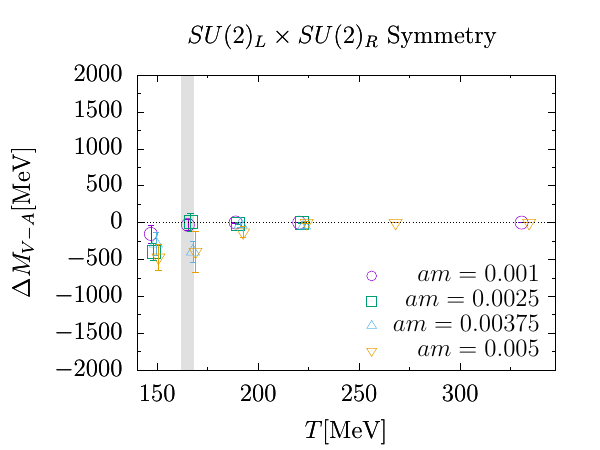}
\caption{The screening mass difference plot for $\csg$ with respect to temperature for the range of masses in this study. \label{fig:Vx-Axnf2sym}}
\end{figure}

\begin{figure}[hb]
\centering
\includegraphics[scale=1.2]{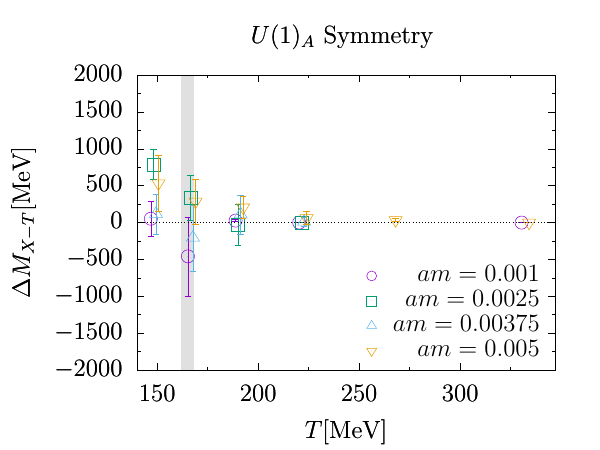}
\caption{The screening mass difference plot for $U(1)_{A}$ with respect to temperature, the axial anomaly appears to be suppressed at $T=189\mv$. \label{fig:Xt-Ttnf2sym}}
\end{figure}

\section{Summary of Preliminary Results}
\label{sec:Conclusion}
Figure \ref{fig:mch-nf2-0010} shows, for the lowest temperature $\sim 0.9T_c$ in the study, a good overlap between screening masses in various channels and the experimental data at $T=0$; in particular, the pseudoscalar overlaps with the $T=0$ $\pi^\pm$ mass indicating that the chiral condensate rapidly converges to the zero temperature value below $T_c$.\par
The most significant difference between our work and the results by HotQCD, plotted as shaded bands in fig \ref{fig:mch-nf2-0010}, is the behavior of the $S$ channel at low temperatures. In \cite{Bazavov:2019www}, the staggered fermion action employed by HotQCD breaks isospin symmetry such that a two pion decay channel from the scalar triplet is opened; when the isospin symmetry is preserved the triplet scalar cannot decay as this decay channel is forbidden.  Because our simulations are done with M\"obius Domain Wall fermions we do not observe the same $2\pi^{\pm}$ lattice artifact, and instead, we observe a large increase in the screening mass in the region of broken chiral symmetry.\par
The $\csg$ symmetry restoration, seen through the mass difference in the $V$ and $A$ partner channels, shows a clear signal for the all masses but is then most evident at $am=0.0010$. From figure \ref{fig:Vx-Axnf2sym} at $T=147\mv$ the $\csg$ chiral symmetry is broken, and it appears to be restored at the transition point $T_c\sim 165\mv$ and remains zero for all higher temperatures. $U(1)_A$ symmetry is also effectively ``restored'' for $T\sim 165\mv$ from the plot of $X_t-T_t$ fig. \ref{fig:Xt-Ttnf2sym}. While it is still unclear if the transition is exactly at the pseudocritical point at $T=189\mv$ corresponding to $T\sim 1.1T_c$ we see restoration of $U(1)_A$ which is below the $1.2-1.3T_c$ threshold for restoration reported by \cite{Bazavov:2019www,Buchoff:2013nra,Cheng:2010fe,HotQCD:2012vvd,Tomiya:2016jwr,Brandt:2016daq} and is observed in the $PS-S$ screening mass difference as well.\par
\begin{figure}[t]
\centering
\includegraphics[scale=1.2]{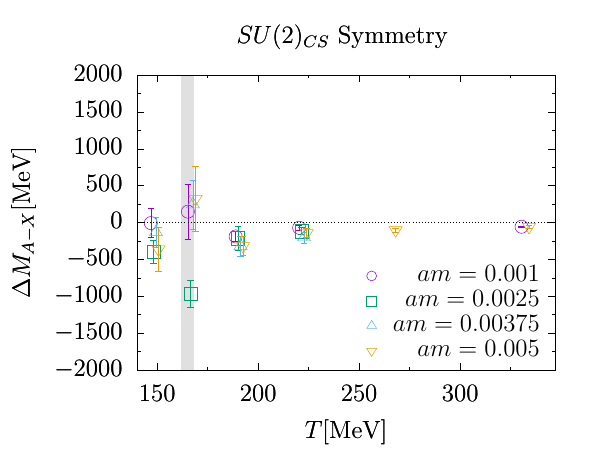}
\caption{The screening mass difference plot for $\CS$ for various temperatures and masses, while it looks close to being restored at $T=330\mv$ it is $\sim 40\mv$ and most likely approximate if the symmetry is emergent at this temperature. \label{fig:Xt-Axnf2sym}}
\end{figure}

In contrast to both of these symmetries we also plot a temperature spectrum for the $\CS$ predicted in \cite{Glozman:2017dfd, Glozman:2015qzf, Glozman:2016swy}, unlike the clear signals of symmetry restoration seen in figs \ref{fig:Vx-Axnf2sym} and \ref{fig:Xt-Ttnf2sym}, the plot of the symmetry pair $X-A$ figure \ref{fig:Xt-Axnf2sym} does not converge to zero for the temperatures in our study. While there is a significant reduction in noise, we do not see a convergence to, or sustained, zero value indicating emergence of $\CS$.

\acknowledgments{
We thank O.~Phillipsen, and R.~Pisarski in particular for their discussions of our work. Discussion of this work in its later stages during the HHIQCD 2024 long term workshop at the Yuakawa Institute for Theoretical Physics was important to several of the revisions of our work in preparation of these proceedings. For the numerical simulation we use the QCD software packages Grid \cite{Boyle:2015tjk,Meyer:2021uoj} for configuration generations, Bridge++ \cite{Ueda:2014rya,Akahoshi:2021gvk} for measurements and the Japan Lattice Data Grid (JLDG) for storing portions of this work. Numerical simulations were performed on Wisteria/BDEC-01 Odyssey at JCAHPC under a support of the HPCI System Research Projects (Project IDs: hp170061, hp180061, hp190090, hp200086, and hp210104) and Fugaku computer provided by the RIKEN Center for Computational Science with support provided by HPCI System Research Projects (Project ID: hp200130, hp210165, hp210231, hp220279, hp230323). This work is supported in part by the Japanese Grant-in-Aid for Scientific Research (No. JP26247043, JP18H01216, JP18H04484,JP22H01219) and by Joint Institute for Computational Fundamental Science (JICFuS). We would also like to acknowledge support from the Deutsche Forshungsgemeinschaft(DFG, German Research Foundation) through the CRC-TR 211 "Strong-interaction matter under extreme conditions" - project number 315477589 - TRR 211.}

\bibliographystyle{./JHEP}
\bibliography{skele-sym}

\providecommand{\href}[2]{#2}\begingroup\raggedright\begin{thebibliography}{10}

\bibitem{Aoki:2012yj}
S.~Aoki, H.~Fukaya and Y.~Taniguchi, \emph{{Chiral symmetry restoration,
  eigenvalue density of Dirac operator and axial U(1) anomaly at finite
  temperature}}, \href{https://doi.org/10.1103/PhysRevD.86.114512}{\emph{Phys.
  Rev. D} {\bfseries 86} (2012) 114512}
  [\href{https://arxiv.org/abs/1209.2061}{{\ttfamily 1209.2061}}].

\bibitem{Aoki:2024uvl}
{\scshape JLQCD:} collaboration, \emph{{Chiral susceptibility and axial U(1)
  anomaly near the (pseudo-)critical temperature}},
  \href{https://doi.org/10.22323/1.453.0184}{\emph{PoS} {\bfseries LATTICE2023}
  (2024) 184} [\href{https://arxiv.org/abs/2401.06459}{{\ttfamily
  2401.06459}}].

\bibitem{Cossu:2013uua}
G.~Cossu, S.~Aoki, H.~Fukaya, S.~Hashimoto, T.~Kaneko, H.~Matsufuru et~al.,
  \emph{{Finite temperature study of the axial U(1) symmetry on the lattice
  with overlap fermion formulation}},
  \href{https://doi.org/10.1103/PhysRevD.87.114514}{\emph{Phys. Rev. D}
  {\bfseries 87} (2013) 114514}
  [\href{https://arxiv.org/abs/1304.6145}{{\ttfamily 1304.6145}}].

\bibitem{Pisarski:1983ms}
R.D.~Pisarski and F.~Wilczek, \emph{{Remarks on the Chiral Phase Transition in
  Chromodynamics}}, \href{https://doi.org/10.1103/PhysRevD.29.338}{\emph{Phys.
  Rev. D} {\bfseries 29} (1984) 338}.

\bibitem{Suzuki:2019vzy}
{\scshape JLQCD} collaboration, \emph{{Axial U(1) symmetry, topology, and Dirac
  spectra at high temperature in $N_f=2$ lattice QCD}},
  \href{https://doi.org/10.22323/1.317.0085}{\emph{PoS} {\bfseries CD2018}
  (2019) 085} [\href{https://arxiv.org/abs/1908.11684}{{\ttfamily
  1908.11684}}].

\bibitem{Rohrhofer:2019yko}
C.~Rohrhofer, Y.~Aoki, G.~Cossu, H.~Fukaya, C.~Gattringer, L.Y.~Glozman et~al.,
  \emph{{Symmetries of the Light Hadron Spectrum in High Temperature QCD}},
  \href{https://doi.org/10.22323/1.363.0227}{\emph{PoS} {\bfseries LATTICE2019}
  (2020) 227} [\href{https://arxiv.org/abs/1912.00678}{{\ttfamily
  1912.00678}}].

\bibitem{Rohrhofer:2019qwq}
C.~Rohrhofer, Y.~Aoki, G.~Cossu, H.~Fukaya, C.~Gattringer, L.Y.~Glozman et~al.,
  \emph{{Symmetries of spatial meson correlators in high temperature QCD}},
  \href{https://arxiv.org/abs/1902.03191}{{\ttfamily 1902.03191}}.

\bibitem{Aoki:2020noz}
{\scshape JLQCD} collaboration, \emph{{Study of the axial $U(1)$ anomaly at
  high temperature with lattice chiral fermions}},
  \href{https://doi.org/10.1103/PhysRevD.103.074506}{\emph{Phys. Rev. D}
  {\bfseries 103} (2021) 074506}
  [\href{https://arxiv.org/abs/2011.01499}{{\ttfamily 2011.01499}}].

\bibitem{Glozman:2015qzf}
L.Y.~Glozman, \emph{{$SU(2N_F)$ hidden symmetry of QCD}},
  \href{https://arxiv.org/abs/1511.05857}{{\ttfamily 1511.05857}}.

\bibitem{Glozman:2017dfd}
L.Y.~Glozman, \emph{{Chiralspin symmetry and QCD at high temperature}},
  \href{https://doi.org/10.1140/epja/i2018-12560-0}{\emph{Eur. Phys. J. A}
  {\bfseries 54} (2018) 117}
  [\href{https://arxiv.org/abs/1712.05168}{{\ttfamily 1712.05168}}].

\bibitem{Glozman:2022lda}
L.Y.~Glozman, O.~Philipsen and R.D.~Pisarski, \emph{{Chiral spin symmetry and
  the QCD phase diagram}},
  \href{https://doi.org/10.1140/epja/s10050-022-00895-4}{\emph{Eur. Phys. J. A}
  {\bfseries 58} (2022) 247}
  [\href{https://arxiv.org/abs/2204.05083}{{\ttfamily 2204.05083}}].

\bibitem{Glozman:2016swy}
L.Y.~Glozman, \emph{{$SU(2N_F)$ symmetry of QCD at high temperature and its
  implications}},
  \href{https://doi.org/10.5506/APhysPolBSupp.10.583}{\emph{Acta Phys. Polon.
  Supp.} {\bfseries 10} (2017) 583}
  [\href{https://arxiv.org/abs/1610.00275}{{\ttfamily 1610.00275}}].

\bibitem{Rohrhofer:2019saj}
C.~Rohrhofer, Y.~Aoki, G.~Cossu, H.~Fukaya, L.~Glozman, S.~Hashimoto et~al.,
  \emph{{Observation of approximate SU(2)$_{CS}$ and SU(2$n_f$) symmetries in
  high temperature lattice QCD}},
  \href{https://doi.org/10.1016/j.nuclphysa.2018.10.037}{\emph{Nucl. Phys. A}
  {\bfseries 982} (2019) 207}.

\bibitem{Rohrhofer:2018xpn}
C.~Rohrhofer, \emph{{Symmetries of QCD at high temperature}}, Ph.D. thesis,
  Graz U., 2018.

\bibitem{Suzuki:2020rla}
{\scshape JLQCD} collaboration, \emph{{Axial U(1) symmetry and mesonic
  correlators at high temperature in $N_f=2$ lattice QCD}},
  \href{https://doi.org/10.22323/1.363.0178}{\emph{PoS} {\bfseries LATTICE2019}
  (2020) 178} [\href{https://arxiv.org/abs/2001.07962}{{\ttfamily
  2001.07962}}].

\bibitem{Zhang:2022kzb}
Y.~Zhang, Y.~Aoki, S.~Hashimoto, I.~Kanamori, T.~Kaneko and Y.~Nakamura,
  \emph{{Finite temperature QCD phase transition with 3 flavors of Mobius
  domain wall fermions}}, \href{https://doi.org/10.22323/1.430.0197}{\emph{PoS}
  {\bfseries LATTICE2022} (2023) 197}
  [\href{https://arxiv.org/abs/2212.10021}{{\ttfamily 2212.10021}}].

\bibitem{Ward:2024tdm}
D.~Ward, S.~Aoki, Y.~Aoki, H.~Fukaya, S.~Hashimoto, I.~Kanamori et~al.,
  \emph{{Study of Chiral Symmetry and $U(1)_A$ using Spatial Correlators for
  $N_f$ = 2 + 1 QCD at finite temperature with Domain Wall Fermions}},
  \href{https://doi.org/10.22323/1.453.0182}{\emph{PoS} {\bfseries LATTICE2023}
  (2024) 182} [\href{https://arxiv.org/abs/2401.07514}{{\ttfamily
  2401.07514}}].

\bibitem{Bala:2023iqu}
D.~Bala, O.~Kaczmarek, P.~Lowdon, O.~Philipsen and T.~Ueding,
  \emph{{Pseudo-scalar meson spectral properties in the chiral crossover region
  of QCD}},  \href{https://arxiv.org/abs/2310.13476}{{\ttfamily 2310.13476}}.

\bibitem{Chiu:2023hnm}
T.-W.~Chiu, \emph{{Symmetries of meson correlators in high-temperature QCD with
  physical (u/d,s,c) domain-wall quarks}},
  \href{https://doi.org/10.1103/PhysRevD.107.114501}{\emph{Phys. Rev. D}
  {\bfseries 107} (2023) 114501}
  [\href{https://arxiv.org/abs/2302.06073}{{\ttfamily 2302.06073}}].

\bibitem{Chiu:2024jyz}
T.-W.~Chiu, \emph{{Symmetries of spatial correlators of light and heavy mesons
  in high temperature lattice QCD}},
  \href{https://doi.org/10.1103/PhysRevD.110.014502}{\emph{Phys. Rev. D}
  {\bfseries 110} (2024) 014502}
  [\href{https://arxiv.org/abs/2404.15932}{{\ttfamily 2404.15932}}].

\bibitem{DallaBrida:2021ddx}
M.~Dalla~Brida, L.~Giusti, T.~Harris, D.~Laudicina and M.~Pepe,
  \emph{{Non-perturbative thermal QCD at all temperatures: the case of mesonic
  screening masses}},
  \href{https://doi.org/10.1007/JHEP04(2022)034}{\emph{JHEP} {\bfseries 04}
  (2022) 034} [\href{https://arxiv.org/abs/2112.05427}{{\ttfamily
  2112.05427}}].

\bibitem{Bazavov:2019www}
A.~Bazavov et~al., \emph{{Meson screening masses in (2+1)-flavor QCD}},
  \href{https://doi.org/10.1103/PhysRevD.100.094510}{\emph{Phys. Rev. D}
  {\bfseries 100} (2019) 094510}
  [\href{https://arxiv.org/abs/1908.09552}{{\ttfamily 1908.09552}}].

\bibitem{Buchoff:2013nra}
M.I.~Buchoff et~al., \emph{{QCD chiral transition, U(1)A symmetry and the dirac
  spectrum using domain wall fermions}},
  \href{https://doi.org/10.1103/PhysRevD.89.054514}{\emph{Phys. Rev. D}
  {\bfseries 89} (2014) 054514}
  [\href{https://arxiv.org/abs/1309.4149}{{\ttfamily 1309.4149}}].

\bibitem{Cheng:2010fe}
M.~Cheng et~al., \emph{{Meson screening masses from lattice QCD with two light
  and the strange quark}},
  \href{https://doi.org/10.1140/epjc/s10052-011-1564-y}{\emph{Eur. Phys. J. C}
  {\bfseries 71} (2011) 1564}
  [\href{https://arxiv.org/abs/1010.1216}{{\ttfamily 1010.1216}}].

\bibitem{HotQCD:2012vvd}
{\scshape HotQCD} collaboration, \emph{{The chiral transition and $U(1)_A$
  symmetry restoration from lattice QCD using Domain Wall Fermions}},
  \href{https://doi.org/10.1103/PhysRevD.86.094503}{\emph{Phys. Rev. D}
  {\bfseries 86} (2012) 094503}
  [\href{https://arxiv.org/abs/1205.3535}{{\ttfamily 1205.3535}}].

\bibitem{Tomiya:2016jwr}
A.~Tomiya, G.~Cossu, S.~Aoki, H.~Fukaya, S.~Hashimoto, T.~Kaneko et~al.,
  \emph{{Evidence of effective axial U(1) symmetry restoration at high
  temperature QCD}},
  \href{https://doi.org/10.1103/PhysRevD.96.034509}{\emph{Phys. Rev. D}
  {\bfseries 96} (2017) 034509}
  [\href{https://arxiv.org/abs/1612.01908}{{\ttfamily 1612.01908}}].

\bibitem{Brandt:2016daq}
B.B.~Brandt, A.~Francis, H.B.~Meyer, O.~Philipsen, D.~Robaina and H.~Wittig,
  \emph{{On the strength of the $U_A(1)$ anomaly at the chiral phase transition
  in $N_f=2$ QCD}}, \href{https://doi.org/10.1007/JHEP12(2016)158}{\emph{JHEP}
  {\bfseries 12} (2016) 158}
  [\href{https://arxiv.org/abs/1608.06882}{{\ttfamily 1608.06882}}].

\bibitem{Boyle:2015tjk}
P.~Boyle, A.~Yamaguchi, G.~Cossu and A.~Portelli, \emph{{Grid: A next
  generation data parallel C++ QCD library}},
  \href{https://arxiv.org/abs/1512.03487}{{\ttfamily 1512.03487}}.

\bibitem{Meyer:2021uoj}
N.~Meyer, P.~Georg, S.~Solbrig and T.~Wettig, \emph{{Grid on QPACE 4}},
  \href{https://doi.org/10.22323/1.396.0068}{\emph{PoS} {\bfseries LATTICE2021}
  (2022) 068} [\href{https://arxiv.org/abs/2112.01852}{{\ttfamily
  2112.01852}}].

\bibitem{Ueda:2014rya}
S.~Ueda, S.~Aoki, T.~Aoyama, K.~Kanaya, H.~Matsufuru, S.~Motoki et~al.,
  \emph{{Development of an object oriented lattice QCD code 'Bridge++'}},
  \href{https://doi.org/10.1088/1742-6596/523/1/012046}{\emph{J. Phys. Conf.
  Ser.} {\bfseries 523} (2014) 012046}.

\bibitem{Akahoshi:2021gvk}
Y.~Akahoshi, S.~Aoki, T.~Aoyama, I.~Kanamori, K.~Kanaya, H.~Matsufuru et~al.,
  \emph{{General purpose lattice QCD code set Bridge++ 2.0 for high performance
  computing}}, \href{https://doi.org/10.1088/1742-6596/2207/1/012053}{\emph{J.
  Phys. Conf. Ser.} {\bfseries 2207} (2022) 012053}
  [\href{https://arxiv.org/abs/2111.04457}{{\ttfamily 2111.04457}}].

\end{thebibliography}\endgroup

\end{document}